# Generation of Electron Whistler Waves at the Mirror Mode Magnetic Holes: MMS Observations and PIC Simulation

N. Ahmadi[1], F. D. Wilder[1], R. E. Ergun[1,2], M. Argall[3], M. E. Usanova[1], H. Breuillard[4], D. Malaspina[1], K. Paulson[5], K. Germaschewski[3,5], S. Eriksson[1], K. Goodrich[1], R. Torbert[3,5], O. Le Contel[4], R. J. Strangeway[6], C. T. Russell[6], J. Burch[7] and B. Giles[8]

[1]Laboratory for Atmospheric and Space Physics, University of Colorado, Boulder, Colorado, USA.

[2]Department of Astrophysical and Planetary Sciences, University of Colorado, Boulder, Colorado, USA.

[3]Space Science Center, University of New Hampshire, Durham, New Hampshire, USA.

[4]Laboratoire de Physique des Plasmas (LPP), UMR7648, CNRS/Ecole Polytechnique/Sorbonne Université/Univ. Paris Sud/Observatoire de Paris, Paris, France.

[5]Department of Physics, University of New Hampshire, Durham, New Hampshire, USA.

[6]University of California, Los Angeles, CA, USA.

[7]Southwest Research Institute, San Antonio, TX, USA.

[8]NASA Goddard Space Flight Center, Greenbelt, MD, USA.

Corresponding author: Narges Ahmadi (Narges.Ahmadi@colorado.edu)

Key Points:

- Electron whistler waves are generated at mirror mode magnetic holes in the magnetosheath.

- Electron temperature anisotropy is enhanced at the center of magnetic holes by trapping and at the edges by Fermi acceleration or deceleration of electrons.

- In deep magnetic holes, low energy electrons resonate with the electron whistler waves and are isotropized while higher energy electrons remain anisotropic inside the magnetic holes.

**Abstract**

The Magnetospheric Multiscale (MMS) mission has observed electron whistler waves at the center and at the edges of magnetic holes in the dayside magnetosheath. The magnetic holes are nonlinear mirror structures since their magnitude is anti-correlated with particle density. In this article, we examine the growth mechanisms of these whistler waves and their interaction with the host magnetic hole. In the observations, as magnetic holes develop and get deeper, an electron population gets trapped and develops a temperature anisotropy favorable for whistler waves to be generated. In addition, the decrease in magnetic field magnitude and the increase in density reduces the electron resonance energy, which promotes the electron cyclotron resonance. To investigate this process, we used an expanding box particle-in-cell simulations to produce the mirror instability, which then evolves into magnetic holes. The simulation shows that whistler waves can be generated at the center and edges of magnetic holes, which reproduces the primary features of the MMS observations. The simulation shows that the electron temperature anisotropy develops in the center of the magnetic hole once the mirror instability reaches its nonlinear stage of evolution. The plasma is then unstable to whistler waves at the minimum of the magnetic field structures. In the saturation regime of mirror instability, when magnetic holes are developed, the electron temperature anisotropy appears at the edges of the holes and electron distributions become more isotropic at the magnetic field minimum. At the edges, the expansion of magnetic holes decelerates the electrons which leads to temperature anisotropies.

# 1 Introduction

Satellite observations frequently find magnetic holes in the planetary (Earth, Saturn and Jupiter) magnetosheaths (*Erdős and Balogh*, 1996; *Bavassano-Cattaneo et al*, 1998; *Soucek et al*, 2008). Magnetic holes are depressions in the background magnetic field formed on a spatial scale greater than the ion gyroradius with the perturbed magnetic field $\delta \boldsymbol{B}$ amplitudes comparable to the background $\boldsymbol{B}$. These structures are believed to be consequence of the nonlinear evolution of mirror instability perturbations (*Erdős and Balogh*, 1996; *Sahraoui et al.,* 2004; *Soucek et al*, 2008; *Califano et al*, 2008; *Génot et al*, 2009; *Ahmadi et al*., 2017). The quasi-perpendicular bow shock heats ions in the direction perpendicular to the background magnetic field, creating a temperature anisotropy $(T_{p\perp} > T_{p\parallel})$ that generates mirror mode structures in the downstream region. $T_{p\perp}$ and $T_{p\parallel}$ stand for proton temperatures perpendicular and parallel to the ambient magnetic field, respectively. Mirror modes form as compressional magnetic peaks or magnetic holes along the background magnetic field and have zero group velocity (they are stationary in the plasma rest frame). As they convect with the plasma toward the magnetopause, in the middle of the magnetosheath (between bow shock and magnetopause) where the plasma is mirror unstable, they form quasi-periodic structures with magnetic peaks being the dominant structures. Close to the magnetopause, the plasma becomes marginally mirror stable causing the magnetic peaks to collapse while magnetic holes survive this transition, as stated by the bi-stability theory (*Kuznetsov et al.*, 2007; *Califano et al.*, 2008). According to the bi-stability theory, magnetic peaks and magnetic holes are the solutions of the Vlasov-Maxwell equations in a mirror unstable plasma but in a mirror stable plasma, only magnetic holes can survive and magnetic peaks collapse.

Soucek et al. (2008) have shown the shape of mirror mode structures depends on plasma parameters. They performed a statistical study of mirror mode magnetic peaks and magnetic holes and their relation to mirror instability threshold using Cluster observations. They showed that magnetic peaks are present when plasma is mirror unstable and magnetic holes dominate in mirror stable or marginally mirror stable plasma. The same behavior for mirror mode waves was observed by Voyager spacecraft on the dayside of Saturn in its magnetosheath as reported by Bavassano-Cattaneo et al. (1998). Ahmadi et al. (2017) developed an expanding and compressing box Particle-in-Cell (PIC) code using the Plasma Simulation Code (PSC) (*Germaschewski et al.*, 2016) in order to model the convection of the mirror mode waves in the dynamic magnetosheath. They showed that mirror mode structures evolve into deep magnetic holes in a marginally mirror stable plasma, in agreement with the Cluster observations. These ion scale magnetic holes can impact the electron dynamics and lead to the generation of electron whistler waves.

Observations also show that mirror mode structures are often accompanied by electromagnetic whistler waves (historically known as "lion roars") in the mirror mode magnetic holes (*Smith and Tsurutani*, 1976; *Tsurutani et al.*, 1982; *Lee et al.*, 1987; *Yearby et al.*, 2005; *Masood et al.*, 2006; *Zhang et al.*, 1998). Electromagnetic whistler waves are generated either by electron temperature anisotropy ($T_{e\perp} > T_{e\parallel}$) or through electron beam-driven instability (*Trakhtengerts et al.*, 1999; *Gary and Karimabadi*, 2006) and propagate mostly parallel to the background magnetic field with frequencies between $\Omega_p < \omega < \Omega_e$, where $\Omega_p$ and $\Omega_e$ are the proton and electron gyrofrequencies, respectively. Gary and Wang (1996) provided an analytical instability threshold for whistler anisotropy instability (hereafter referred to as whistler waves) in a bi-Maxwellian plasma with maximum growth rate $\gamma_m = 0.01\Omega_e$:

$$R_w = \beta_{e\parallel}^{0.55} \left(\frac{T_{e\perp}}{T_{e\parallel}} - 1\right) - 0.36 \qquad (1)$$

A plasma is whistler unstable when $R_w \geq 0$ and whistler stable when $R_w < 0$.

Mirror mode waves play a dominant role in mitigating the proton temperature anisotropy in the magnetosheath, but they do not fully isotropize the magnetosheath plasma; it remains in a constant state of marginal stability. Recent multi-satellite missions using spacecraft at ion- and electron-scale separations, improved precision, and faster sampling rates, as well as advances in numerical computing, have made it possible to study cross-scale energy transport, making this study a timely endeavor to understand transport of energy from ion-scale structures to the electrons.

Presence of electron temperature anisotropy can also enhance mirror mode growth rate but mirror instability has to compete with the whistler instability for consuming the available free energy. We previously showed that the whistler instability quickly grows and saturates, thereby consuming most of the electron free energy as the mirror instability is just starting to grow (*Ahmadi et al.*, 2016). Previous results have shown that electrons require a generator of free energy in order to influence the growth of mirror mode waves (*Pantellini and Schwartz,* 1995; *Pokhotelov et al.,* 2000; *Ahmadi et al.*, 2016), and that plasma expansion and field line draping near the magnetopause can provide the energy source.

Whistler waves have been observed in the center and at the edges of the mirror mode holes (*Smith and Tsurutani*, 1976; *Zhang et al.*, 1998; *Tenerani et al.,* 2013; *Tenerani et al.,* 2014). Several mechanisms have been previously proposed for the generation of these whistler waves (*Thorne and Tsurutani*, 1981; *Zhima et al.*, 2015), but the electron dynamics in magnetosheath mirror modes are not yet fully understood. Early theories hypothesize that the free energy for whistler wave growth is provided by the pitch angle anisotropy of low-energy magnetosheath electrons. As particles with large pitch angle (close to 90) become trapped by the growing magnetic holes, the mirror force enhances anisotropy in the low-field region (*Lee at al.*, 1987). Density increases as the field weakens, reducing the scaling energy ($E_M = B^2/8\pi N$) for resonant wave-particle interactions to values as low as $10 eV$ where $N$ is the particle density (*Thorn and Tsurutani*, 1981). The result is an exponential increase in the number of electrons that are capable of resonating with the waves. These theories, however, ignore the forces that develop as the mirror structure grows. The nonlinear growth phase of the mirror instability introduces forces that alter the simple trapping model. The trapped plasma with turning points at higher $|B|$ is Fermi accelerated in the parallel direction as the mirror mode peaks grow. Meanwhile, the trapped plasma deep within the mirror mode holes is cooled in both the perpendicular and parallel directions by Fermi and betatron deceleration as the troughs deepen (*Kivelson and Southwood*, 1996). Such processes are confirmed by spacecraft observations of electron distribution functions in the peaks and troughs of mirror modes (*Leckband et al.*, 1995; *Chisham et al.*, 1998; *Soucek and Escoubet*, 2011; *Breuillard et al*, 2017). Another counterpoint to the trapping model is that whistler waves have been observed on the edges of magnetic holes (*Zhima et al.*, 2015). Also, Ahmadi et al. (2017) have observed signatures of whistler waves at the minimum and the edges of mirror mode holes in PIC expanding box simulations, which we discuss further in this paper. Zhang et al., (1998) showed that 30% of magnetosheath whistler waves are observed within mirror mode structures, meaning there is a driver of free energy. The dominant frequencies of these whistler waves are around $0.12\Omega_e$.

More recent observations of mirror modes in the magnetotail report both cigar-shaped and pancake-shaped distributions within the holes, and suggest that different evolutionary phases of mirror modes will produce different shapes in the electron distribution functions (*Li et al.*, 2014). Regardless of the shape, whistler waves were present.

In this paper, we investigate how electron whistler waves are generated at the center and edges of ion-scale magnetic holes in the magnetosheath and study the electron distributions inside the holes using high resolution Magnetospheric Multiscale (MMS) mission data and PIC simulations. In section 2, we review our expanding box PIC simulations and show that whistler waves are able to grow in the mirror mode magnetic holes. Section 3 shows MMS observations of whistler waves inside mirror mode magnetic holes in the magnetosheath. In section 4, we investigate the driver of electron whistler waves and provide details of different mechanism for electron scattering. And finally, we summarize our results in section 5.

**2 Generation of Electron Whistler Waves in PIC Simulations**

We performed two-dimensional PIC expanding box simulation to produce the mirror mode magnetic holes using PSC (*Germaschewski et al.*, 2016; *Ahmadi et al.*, 2017). The expanding box method implemented into PSC is described by Sironi and Narayan (2016). We expand the simulation box in the direction of the ambient magnetic field $\boldsymbol{B} = \boldsymbol{B_0}\hat{\boldsymbol{z}}$ (parallel expansion). The simulation parameters are given in Ahmadi et al. (2017). The initial temperature anisotropy and parallel plasma beta for protons (index p) and electrons (index e) are $T_{p\perp}/T_{p\parallel} = 1.1$, $T_{e\perp}/T_{e\parallel} = 1.$, $\boldsymbol{\beta_{p\parallel}} = 13$ and $\boldsymbol{\beta_{e\parallel}} = 1$. $\boldsymbol{\beta}$ is the ratio of particle pressure to magnetic pressure. The simulation starts with Maxwellian electrons and slightly anisotropic protons. If there are no waves present in the system, plasma follows the adiabatic path according to Chew-Goldberger-Low (CGL) model (*Chew et al.*, 1956). In a parallel expansion, the conservation of first and second adiabatic invariants increases the temperature anisotropies and lead to generation of proton cyclotron and mirror instabilities. We start the simulation with a small electron beta ($\boldsymbol{\beta}_{e\parallel} = \boldsymbol{1}$) and it decreases as we expand the simulation box. The decrease in plasma beta resembles the magnetosheath plasma beta from the bow shock to the magnetopause. The plasma is stable to the whistler instability when we average the plasma parameters in the entire simulation domain according to analytical threshold in equation (1). The expansion leads to high averaged electron temperature anisotropy as shown in Figure 1 while electron beta is decreasing. At the nonlinear regime of mirror instability when magnetic holes are shallow, the proton and electron temperature anisotropy are anti-correlated with the magnetic field magnitude in the magnetic hole. Although the averaged plasma parameters are electron whistler stable with $\boldsymbol{R_w < 0}$, which is shown in Figure 1, the plasma is locally unstable to the whistler waves at the minimum of the magnetic field structures as displayed in Figure 2a. In the saturation regime of mirror instability, when magnetic holes are dominant and deep, electrons become isotropic at the magnetic field minimum while electron temperature anisotropy still exists at the edges of the magnetic holes (Figure 2b). This can be explained by the generation of whistler waves at the magnetic field minimum. Electrons are isotropized at the minimum and anisotropy remains at the edges of the magnetic hole. The presence of electron temperature anisotropy at the edges of magnetic hole can lead to the generation of the whistler instability at the edges which is seen in MMS observations.

**3 MMS Observations**

The MMS spacecraft has spent a large amount of time in the magnetosheath and has crossed the magnetopause many times to capture the dynamics of the magnetic reconnection (*Burch et al.,* 2016) and it has provided many observations of mirror mode structures in the magnetosheath. One of the unique features of MMS mission is the high time resolution of particle and field measurements when the spacecraft is in burst mode (*Le Contel et al., 2014; Lindqvist et al.*, 2014; *Russell et al.*, 2014; *Baker et al.*, 2016; *Fuselier et al.*, 2016; *Torbert et al.*, 2016; *Pollock et al.*, 2016; *Ergun et al.,* 2016). Using burst mode MMS data, there have been recent studies of whistler waves in the magnetosheath associated with mirror mode structures (*Breuillard et al*, 2017). MMS satellites crossed the magnetosheath at approximately 16:20 UT on 1 December 2016. Figure 3 shows an overview of the magnetosheath parameters by MMS1 between 16:20 and 16:23 UT. The first and second panels show ion and electron energy fluxes provided by Fast Plasma Investigation (FPI) instrument in burst resolution 150ms for ions and 30ms for electrons. The third panel shows the magnetic field components and total magnetic field. The background magnetic field is about 45nT with the magnetic field vector being predominantly in the GSE Z direction. The minimums in magnetic field are about 20nT below the background field. These are mirror mode magnetic holes and the magnitude of the magnetic field is anti-correlated with ion density, shown in panel 4. These structures are on the order of the ion gyroradius and they are seen on all spacecraft. The ion and electron temperature anisotropies are shown in panel 5 with red and blue lines, respectively. The ions are anisotropic with $T_{p\perp}/T_{p\parallel} > 1$ and electrons are relatively isotropic except in the magnetic holes. Figure 4 displays the wave polarization analysis of this mirror mode event. Panels 1 and 2 in Figure 4 show magnetic and electric field power spectral density, respectively, which displays electromagnetic whistler wave signatures inside the mirror mode magnetic holes with frequencies below the electron gyrofrequency. Wave polarization analysis using Search Coil Magnetometer data (8192 S/s) and Flux Gate Magnetometer demonstrates that these waves are parallel propagating (Panel 3) and right hand circularly polarized (Panel 4) which confirms they are whistler waves. We use the wavelet (Morlet) polarization analysis in the frequency range $[10 - 2000]$ HZ. These whistler waves are observed at different frequencies depending on the magnetic hole depth. For the shallow magnetic hole at 16:20:34, the frequency of whistlers is about $0.35 f_{ce}$ while for the following deeper holes, the observed whistler frequency is $0.12 f_{ce}$. We observe that as the magnetic hole gets deeper, the whistler waves move to lower frequencies. We also observe the presence of whistler waves at the edges of magnetic holes at 16:21:05, 16:21:24, 16:22:05 and 16:22:40 which agrees with PIC simulation results shown in Figure 2. We see that the wave signatures at the edges of the magnetic holes are at higher frequencies compared to the waves at the minimum of the magnetic structure. We explain the possible generation mechanism of whistler waves inside the magnetic holes in the next section.

**4 Driver of Electron Whistler Waves in Mirror Mode Magnetic Holes**

In the context of mirror modes, there are three mechanisms that can produce unstable distributions to whistler waves: 1) trapping, 2) Fermi acceleration or deceleration, and 3) betatron acceleration. In their non-linear growth phase, the peaks and holes of mirror modes grow in amplitude such that the difference between the maximum and minimum field values of the mirror mode becomes larger (*Kivelson and Southwood*, 1996). The result is that more particles become trapped in the bottle structure. Since the pitch angles of trapped electrons are centered around 90°, the electron plasma becomes more anisotropic and may trigger the whistler resonance condition. Another driver of anisotropy is betatron acceleration. As the magnetic bottles contract or expand on a timescale long enough to conserve the 1st adiabatic invariant $\mu$, the changing magnetic field strength through the plane of the electron gyromotion induces an electric field, causing acceleration or deceleration in the perpendicular direction. Depending on how the mirror structure is evolving and where the turning points are for the trapped electrons, betatron acceleration can either work with or against the trapping mechanism toward the whistler instability threshold. Whistler waves may also be triggered by field-aligned beams that have been generated through Fermi acceleration. Mirror points that approach one another slower than the time it takes electrons to reflect, so that the second adiabatic invariant is conserved, transfer parallel momentum to the electrons that are mirrored. Conversely, mirror points that retreat from one another steal parallel momentum from electrons. If an electron is only weakly trapped, it may receive many parallel energy boosts on its way through the mirror mode system (*Kivelson and Southwood*, 1996; *Drake et al.*, 2006) and become part of an electron beam that destabilizes the trapped plasma.

One final factor that can contribute to plasma destabilization is the resonant energy, which depends on the total field strength. Near the center of the mirror mode holes, where |B| is at a minimum, the resonant energy approaches or drops below the thermal energy of the plasma. This may facilitate wave growth through any of the three generation mechanisms listed above. Figure 5 displays the resonant energy measurements for observed whistler waves. In a resonant instability, resonant particles exchange energy with the wave. For the resonant electrons, the Doppler shifted wave frequency in the frame that moves with the electrons along the magnetic field is an integer ($n$) multiple of the electron's gyrofrequency ($\omega - k_\parallel v_\parallel = n\Omega_e$). The cyclotron resonance energy for $n = 1$ and parallel propagation ($\theta = 0$) is calculated by (*Kennel and Petschek*, 1966)

$$E_R = \frac{B^2}{2\mu_0 N} \frac{\Omega_e}{\omega} \left(1 - \frac{\omega}{\Omega_e}\right)^3$$

In Figure 5a, the dashed black line depicts the resonance energy for $\omega/\Omega_e = 0.35$ which corresponds to the shallow magnetic hole and solid black line shows the resonance energy for deeper magnetic holes with $\omega/\Omega_e = 0.12$. The resonance energy for shallow magnetic hole at 16:20:40 is about $30 eV$ which lies in the center of electron population. The deep magnetic hole's resonance energy is higher and spans the energy range between $40 eV$ and $60 eV$. This confirms that the majority of electron population is cyclotron resonance and is able to generate whistler waves. Figure 5b shows the electron pitch angle distribution for all energies. We see different distribution patterns for the shallow magnetic hole compared to the deep ones. Electrons are concentrated around 90-degree pitch angles at the shallow magnetic hole. At the deep magnetic holes, electrons have a donut shape distribution with electrons being 0 and 180-degree pitch angled at the minimum and 90-degree pitch angled at the edges of the magnetic holes. The averaged electron temperature anisotropy in both cases agree with our PIC simulation in Figure 2. The observed shallow hole corresponds to the young magnetic hole in our simulation with electron temperature anisotropy being maximum at the center of the hole. The observed deep magnetic holes correspond to saturated deep magnetic hole in simulation with electrons being isotropized and scattered at the center of the hole and anisotropic at the edges.

To test these different generation mechanisms in more details, we have plotted the electron pitch angle distributions for different energy ranges in Figure 6. Over-plotted on the pitch angle distributions is the trapping angle for $B_{max} = 45$nT. The trapping angle is given by (*Breuillard et al.,* 2017)

$$\theta_{tr} = sin^{-1}\left(\sqrt{|B|/|B_{max}|}\right)$$

where $B$ is the total magnetic field. Electrons with pitch angles within the plotted black solid lines ($\theta_{tr}$ and $180 - \theta_{tr}$) can get trapped in the magnetic holes. Figure 6 shows different patterns in the pitch angle distribution depending on electron energy range. For energies below 40eV, electrons are moving anti-parallel to the ambient field. In the shallow hole (first magnetic hole) electrons are trapped right within the predicted trapping angles for energies above 40eV. It is interesting that for deep magnetic holes, we see that electrons with 40 to 80eV have donut shape pitch angle distribution. This is the cyclotron resonant population and its pitch angle distribution shows they have been scattered and they are becoming isotropized. Electrons with energies larger than 80eV are mostly still trapped inside deep magnetic holes with their pitch angle close to 90 degrees. This means that the higher energy populations are anisotropic with $T_{e\perp}/T_{e\parallel} > 1$ since electrons are trapped because they are too energetic to interact with the whistler wave.

According to Kivelson and Southwood model for lower energies, the cooling of deeply trapped electrons (~90°) is more effective and for higher energies the heating of shallowly trapped electrons (~close to the trapping angle) is more effective. If we also take into account that electrons interact at the same time with whistlers (thus their pitch-angle is shifted towards parallel values and the electrons can be untrapped) then this is why we observe electrons with 60 to 80eV look isotropic and they may have $T_{e\perp}/T_{e\parallel} < 1$. For energies larger than 80 eV, we also start to see more parallel electrons than anti-parallel, which could be the feature of a nonlinear interaction, as described in Breuillard et al, (2017).

Based on linear dispersion theory, for a known electron temperature anisotropy and plasma beta, the electron whistler instability has a certain frequency and wavelength for a maximum growth rate. A certain group of electrons that have energies close to the resonance energy will resonate with the wave and transfer their perpendicular energy into parallel energy. As the instability grows and relaxes the distribution toward isotropy, the frequency of the excited wave shifts toward lower frequencies and larger wavelength. This causes the resonance energy to transfer to higher energy particles and now they can resonate with the wave and become isotropized. Therefore, the shallow magnetic hole is a young hole that has not reached its nonlinear stage and electrons have not started to interact with either the magnetic structure or the whistlers. The deep magnetic holes are saturated holes since electrons within a certain range of energy (below resonance energy) are already scattered and isotropized.

The observed wave signatures at the edges of the deep magnetic holes correspond to a population of electrons that are 90-degree pitch angled. This leads to an electron temperature anisotropy of larger than one which favors the growth of whistler waves. Figure 6 shows that electrons are perpendicularly pitch angled for all energies higher than 40eV in the deep holes. Fermi deceleration or betatron acceleration of trapped electrons can make them unstable to whistler waves at the edges of the deep magnetic holes.

## 5 Summary and Conclusions

In the present study, we investigated cross scale energy transport between mirror mode waves and electron whistler waves in the magnetosheath. Mirror mode structures are formed downstream of the quasi-perpendicular shock layer and they are convected toward the magnetopause with the magnetosheath flow. Close to the magnetopause, they form magnetic holes which trap electrons. These magnetic holes also develop an electron temperature anisotropy which leads to generation of electron whistler waves. A PIC expanding box simulation showed that electron temperature anisotropy is enhanced at the center and the edges of mirror mode magnetic holes. The expansion and compression of magnetic holes, as they grow, impact the electron population by Fermi acceleration or deceleration and betatron acceleration. Also, as magnetic holes get deeper, whistler waves are observed at lower frequencies. Electron whistler waves observed in the shallow magnetic hole are high frequency with $\boldsymbol{\omega/\Omega_e = 0.35}$ and electrons are 90-degree pitch angled at the center of the hole. An analysis of pitch angle distribution for different energy ranges show that all the energy ranges are 90-degree pitch angled except for distributions with energies below 40eV, which is approximately the cyclotron resonance energy for electrons in the shallow magnetic hole. The rest of the magnetic holes in the observed mirror mode event are deeper. In these holes, the observed frequency of electron whistler waves at the center of the hole is $\boldsymbol{\omega/\Omega_e = 0.12}$, which is lower than for the shallow magnetic hole whistler waves. The pitch angle analysis of different energy ranges show that electrons with energies below 60eV are scattered and they are 0 and 180 pitch angled (field-aligned) at the center of the holes (donut shape distribution) while higher energies are 90-degree pitch angled. The lower energy populations are already isotropized while there is anisotropy in higher energy population. Based on linear dispersion theory and scattering of electron distributions for different energy ranges, we conclude that the shallow magnetic hole is a young hole at the early stage of the evolution while the deep magnetic holes are saturated holes. We also observed signatures of electron whistler waves at the edges of the deep magnetic hole both in simulation and MMS observations. In the observed deep magnetic holes, electrons are 90-degree pitch angled at the edges of the magnetic holes which represents an electron temperature anisotropy favorable for whistler waves to grow. This electron temperature anisotropy can develop due to magnetic hole expansion or contraction which leads to Fermi acceleration or deceleration depending on how deeply the electrons are trapped.

**Acknowledgments and Data**

This work was funded by the NASA MMS project. French involvement (SCM instruments) on MMS is supported by CNES, CNRS-INSIS and CNRS-INSU. The authors recognize the tremendous effort in developing and operating the MMS spacecraft and instruments and sincerely thank all involved. Computations were performed using the following resources: Trillian, a Cray XE6m-200 supercomputer at UNH supported by NSF MRI program under grant PHYS-1229408, and XSEDE resources under contract TG-MCA98N022. MMS spacecraft data is available via the MMS Science Data Center (https://lasp.colorado.edu/mms/sdc/public/). The simulation data is stored on Trillian supercomputer at UNH. Readers interested in attaining the simulation data used should contact the corresponding author at Narges.Ahmadi@colorado.edu.

**References**

Ahmadi, N., Germaschewski, K. & Raeder, J. (2017), *Simulation of observed magnetic holes in the magnetosheath, Physics of Plasmas*, 24, 122121. https://doi.org/10.1063/1.5003017

Ahmadi, N., Germaschewski, K. & Raeder, J. (2016), *Effects of electron temperature anisotropy on proton mirror instability evolution*, Journal of Geophysical Research Space Physics, 121, 5350-5365. doi:10.1002/2016JA022429

Baker, D. N., Riesberg, L., Pankratz, C. K. et al. (2016), *Magnetospheric Multiscale Instrument Suite Operations and Data System*, Space Science Reviews, 199, 545-575. doi:10.1007/s11214-014-0128-5

Bavassano-Cattaneo, M. B., Basile, C., Moreno, G. & Richardson, J. D. (1998), *Evolution of mirror structures in the magnetosheath of Saturn from bow shock to the magnetopause*, Journal of Geophysical Research Space Physics, 103, 11,961-11,972. doi:10.1029/97JA03683

Burch, J. L., Moore, T. E., Torbert, R. B. & Giles, B. L. (2015), *Magnetospheric Multiscale Overview and Science Objectives*, Space Science Reviews, doi:10.1007/s11214-015-0164-9

Breuillard, H., Contel, O. Le., Chust, T., Berthomier, M., Retino, A., Turner, D. L., et al. (2017), *The properties of lion roars and electron dynamics in mirror-mode waves observed by the Magnetospheric Multiscale mission*, Journal of Geophysical Research Space Physics, 122. https://doi.org/10.1002/2017JA024551

Califano, F., Hellinger, P., Kuznetsov, E., Passot, T., Sulem, P. L. & Trávnícek, P. M. (2008), *Nonlinear mirror mode dynamics: Simulations and modeling*, Journal of Geophysical Research Space Physics, 113, A08219. doi:10.1029/2007JA012898

Chew, G. F., Goldberger, M. L. & Low, F. E. (1956), *The Boltzman Equation and the One-Fluid Hydromagnetic Equations in the Absence of Particle Collisions*, Proceedings of the Royal Society A: Mathematical, Physical and Engineering Sciences. doi:10.1098/rspa.1956.0116

Chisham, G., D. Burgess, S. J. Schwartz, and M. W. Dunlop (1998), *Observations of electron distributions in magnetosheath mirror mode waves*, Journal of Geophysical Research Space Physics, 103(A11), 26765–26774, doi:10.1029/98ja02620.

Drake, J. F., Swisdak, M., Che, H., & Shay, M. A. (2006), *Electron acceleration from contracting magnetic islands during reconnection*, Nature, 443(7111), 553-556. http://dx.doi.org/10.1038/nature05116

Erdős, G., and Balogh, A. (1996), *Statistical properties of mirror mode structures observed by Ulysses in the magnetosheath of Jupiter*, Journal of Geophysical Research Space Physics, 101(A1), 1-12. https://doi.org/10.1029/95JA02207

Ergun, R. E., Tucker, S., Westfall, J., Goodrich, K. A., Malaspina, D. M., Summers, D., ... Cully, C. M. (2016), *The Axial Double Probe and Fields Signal Processing for the MMS Mission*, Space Science Reviews, 199(1), 167-188. https://doi.org/10.1007/s11214-014-0115-x

Fuselier, S. A., Lewis, W. S., Schiff, C., Ergun, R., Burch, J. L., Petrinec, S. M., & Trattner, K. J. (2016), *Magnetospheric Multiscale Science Mission Profile and Operations*, Space Science Reviews, 199(1), 77–103. https://doi.org/10.1007/s11214-014-0087-x

Génot, V., E. Budnik, P. Hellinger, T. Passot, G. Belmont, P. M. Trávnícek, P. L. Sulem, E. Lucek, and I. Dandouras (2009), *Mirror structures above and below the linear instability threshold: Cluster observations, fluid model and hybrid simulations*, Annales Geophys, 27, 601-615.

Gary, S. P., J. Wang (1996), *Whistler instability: Electron anisotropy upper band*, Journal of Geophysical Research Space Physics, 101, 10749-10754.

Gary, S. P., H. Karimabadi (2006), *Linear theory of electron temperature anisotropy instabilities: Whistler, mirror, Weibel*, Journal of Geophysical Research Space Physics, 111. doi:10.1029/2006JA011764

Germaschewski, K., W. Fox, S. Abbott, N. Ahmadi, K. Maynard, L. Wang, H. Ruhl, and A. Bhattacharjee (2016), *The plasma simulation code: A modern particle-in-cell code with patch-based load balancing*, Journal of Computational Physics, 318, 305-326.

Kennel, C. F., and H. E. Petschek (1966), *Limit on stably trapped particle fluxes,* Journal of Geophysical Research Space Physics, 71, 1.

Kivelson, M. G., and Southwood, D. J. (1996), *Mirror instability II: The mechanism of nonlinear saturation,* Journal of Geophysical Research Space Physics, 101(A8), 17365–17371. https://doi.org/10.1029/96JA01407

Kuznetsov, E. A., T. Passot, and P. L. Sulem (2007), *Dynamical model for nonlinear mirror modes near threshold*, Physical Review Letters, 98, 235003. doi:10.1103/PhysRevLett.98.235003

Le Contel, O., Leroy, P., Roux, A., Coillot, C., Alison, D., Bouabdellah, A., ... de la Porte, B. (2014), *The Search-Coil Magnetometer for MMS*, Space Science Reviews, *199*(1), 257–282. https://doi.org/10.1007/s11214-014-0096-9

Leckband, J. A., D. Burgess, F. G. E. Pantellini, and S. J. Schwartz (1995), *Ion distribution associated with mirror waves in the Earth's magnetosheath*, Adv. Space. Res. Vol. 15, No. 8/9, pp. (8/9)345-(8/9)348.

Lee, L. C., Wu, C. S., & Price, C. P. (1987), *On the generation of magnetosheath lion roars*, Journal of Geophysical Research Space Physics, *92*(A3), 2343–2348. https://doi.org/10.1029/JA092iA03p02343

Li, H., Zhou, M., Deng, X., Yuan, Z., & Huang, S. (2014), *Electron dynamics and wave activities associated with mirror mode structures in the near-Earth magnetotail*, Science China Technological Sciences, 57(8), 1541–1551. https://doi.org/10.1007/s11431-014- 5574-5

Lindqvist, P.-A., Olsson, G., Torbert, R. B., King, B., Granoff, M., Rau, D., ... Tucker, S. (2014), *The Spin-Plane Double Probe Electric Field Instrument for MMS*, Space Science Reviews, 199(1), 137–165. https://doi.org/10.1007/s11214-014-0116-9

Masood, W., Schwartz, S. J., Maksimovic, M., & Fazakerley, A. N. (2006), *Electron velocity distribution and lion roars in the magnetosheath*, Annales Geophysicae, 24(6), 1725–1735. https://doi.org/10.5194/angeo-24-1725-2006

Pantellini, F. G. E., and S. J. Schwartz (1995), *Electron temperature effects in the linear proton mirror instability*, Journal of Geophysical Research Space Physics, 100, 3539–3549, doi:10.1029/94JA02572.

Pokhotelov, O. A., M. Balikhin, H. S.-C. K. Alleyne, and O. G. Onishchenko (2000), *Mirror instability with finite electron temperature effects*, Journal of Geophysical Research Space Physics, 105, 2393-2402.

Pollock, C., Moore, T., Jacques, A., Burch, J., Gliese, U., Saito, Y., ... Zeuch, M. (2016). Fast Plasma Investigation for Magnetospheric Multiscale, Space Science Reviews, 199(1), 331– 406. https://doi.org/10.1007/s11214-016-0245-4

Russell, C. T., Anderson, B. J., Baumjohann, W., Bromund, K. R., Dearborn, D., Fischer, D., ... Richter, I. (2014), *The Magnetospheric Multiscale Magnetometers*, Space Science Reviews, 199(1), 189–256. https://doi.org/10.1007/s11214-014-0057-3

Sahraoui, F., G. Belmont, J. L. Pinçon, L. Rezeau, A. Balogh, P. Robert and N. Cornilleau-Wehrlin (2004), *Magnetic turbulent spectra in the magnetosheath: new insights*, Annales Geophysicae 22, 2283-2288.

Sironi, L. and R. Narayan (2015), *Electron heating by the ion cyclotron instability in collisionless accretion flows. I. Compression-driven instabilities and the electron heating mechanism*, ApJ, 800, 88.

Smith, E. J., & Tsurutani, B. T. (1976), *Magnetosheath lion roars*, Journal of Geophysical Research, 81(13), 2261–2266. https://doi.org/10.1029/JA081i013p02261

Soucek, J., S. Jan, L. Elizabeth, and D. Iannis (2008), *Properties of magnetosheath mirror modes observed by Cluster and their response to changes in plasma parameters*, Journal of Geophysical Research Space Physics, 113(A4), doi:10.1029/2007ja012649.

Soucek, J., Escoubet, C. P. (2011), *Cluster observations of trapped ions interacting with magnetosheath mirror modes*, Annales Geophysicae, 29(6), 1049-1060.

Tenerani, A., O. Le Contel, F. Califano, P. Robert, D. Fontaine, N. Cornilleau-Wehrlin, and J.-A. Sauvaud (2013), *Cluster observations of whistler waves correlated with ion-scale magnetic structures during the 17 August 2003 substorm event*, Journal of Geophysical Research Space Physics, 118, 6072–6089, doi:10.1002/jgra.50562.

Tenerani, A., O. Le Contel, F. Califano, F. Pegoraro, P. Robert, N. Cornilleau-Wehrlin, and J. A. Sauvaud, *Coupling Between Whistler Waves and Ion-Scale Solitary Waves: Cluster Measurements in the Magnetotail During a Substorm*, Physical Review Letters 109, 155005

Thorne, R. M., and B. T. Tsurutani (1981), *The generation mechanism for magnetosheath lion roars*, Nature, 293(5831), 384–386, doi:10.1038/293384a0.

Torbert, R. B., et al (2016), *The electron drift instrument for MMS,* Space Science Reviews, 199, 283-305. doi:10.1007/s11214-015-0182-7

Trakhtengerts, V. Y., Hobara, Y., Demekhov, A. G., & Hayakawa, M. (1999), *Beam-plasma instability in inhomogeneous magnetic field and second order cyclotron resonance effects*, Physics of Plasmas, 6(3), 692–698. https://doi.org/10.1063/1.873305

Tsurutani, B. T., E. J. Smith, and R. R. Anderson (1982), *Lion roars and nonoscillatory drift mirror waves in the magnetosheath*, Journal of Geophysical Research Space Physics, 87, 6060-6072.

Yearby, K. H., Alleyne, H. S. C., Cornilleau-Wehrlin, N., Santolik, O., Balikhin, M. A., Walker, S. N., ... Lahiff, A. (2005), *Observations of lion roars in the magnetosheath by the STAFF/DWP experiment on the Double Star TC-1 spacecraft*, Annales Geophysicae, 23(8), 2861–2866. https://doi.org/10.5194/angeo-23-2861-2005

Zhang, Y., Matsumoto, H., & Kojima, H. (1998), *Lion roars in the magnetosheath: The Geotail observations,* Journal of Geophysical Research Space Physics, 103(A3), 4615–4626. https://doi.org/10.1029/97JA02519

Zhima, Z., Z. Zeren, C. Jinbin, F. Huishan, L. Wenlong, C. Lunjin, D. Malcolm, X. M. Zhang, and X. H. Shen (2015), *Whistler mode wave generation at the edges of a magnetic dip*, Journal of Geophysical Research Space Physics, 120(4), 2469–2476, doi:10.1002/2014ja020786.

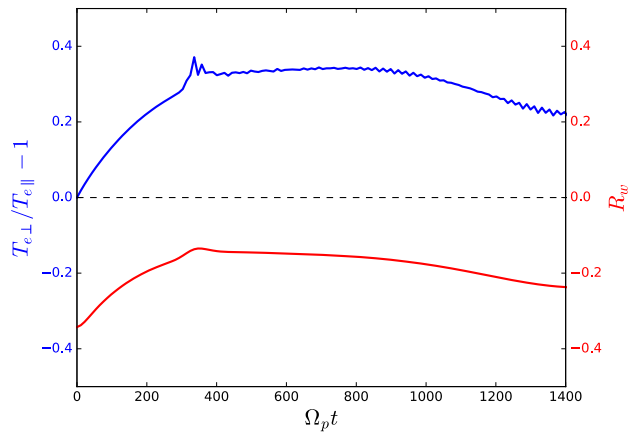

Figure 1: Evolution of average electron temperature anisotropy (blue line) and electron whistler instability threshold (red line) as a function of time in PIC expanding box simulation. Plasma is stable to electron whistler instability ($R_w < 0$), although the average electron temperature anisotropy is high.

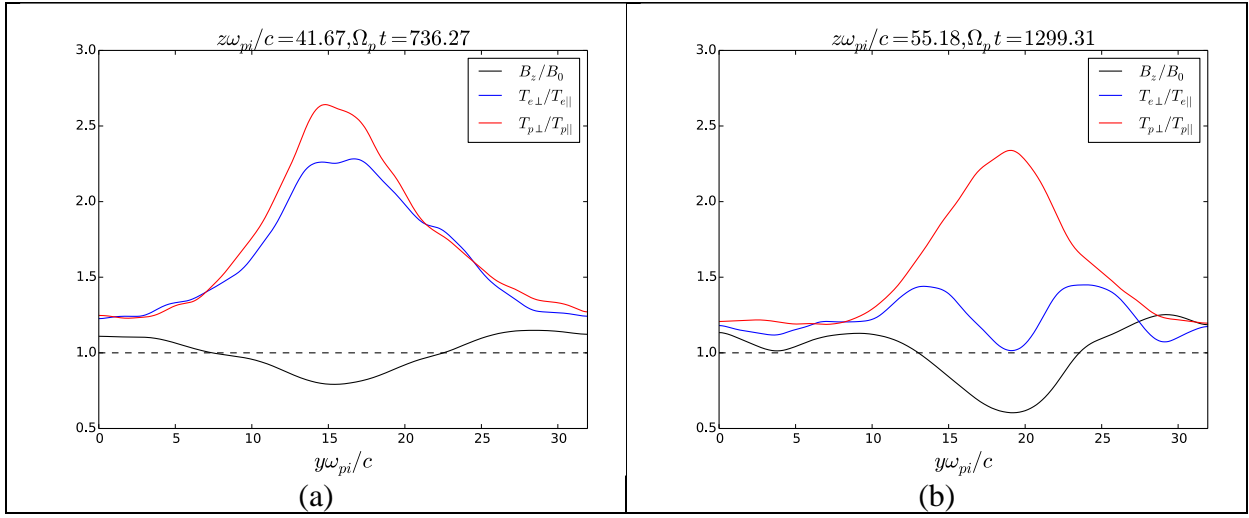

Figure 2: Relation between proton and electron temperature anisotropies and mirror mode magnetic hole. Black line shows the $B_z/B_0$, the red line is proton temperature anisotropy and the blue line represents the electron temperature anisotropy (a) $\Omega_p t = 736$ and (b) $\Omega_p t = 1300$.

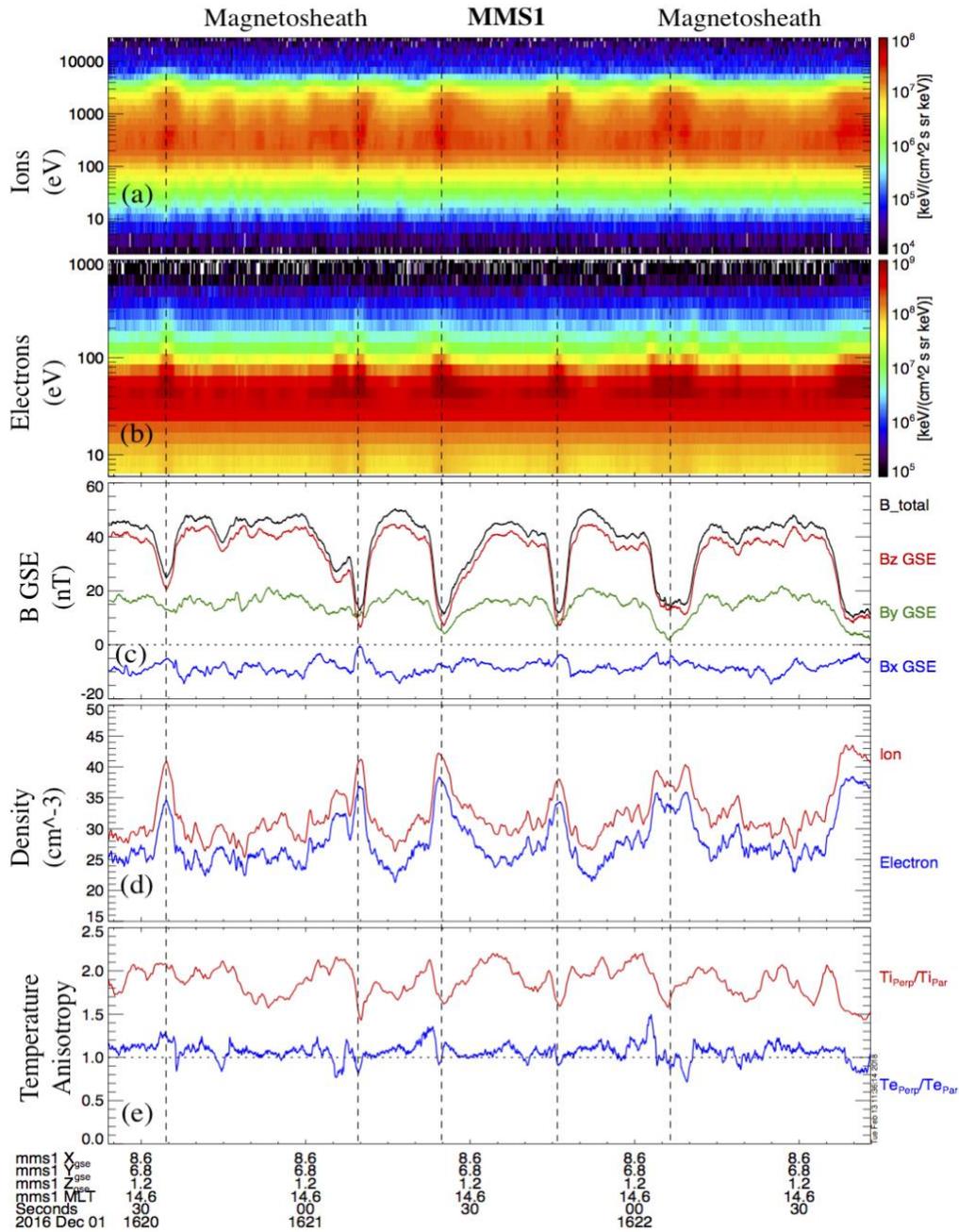

Figure 3: Overview of mirror instability event. (a) Ion energy flux, (b) electron energy flux, (c) magnetic field components and total magnetic field (d) ion (red line) and electron (blue line) number densities, (e) ion (red line) and electron (blue line) temperature anisotropies.

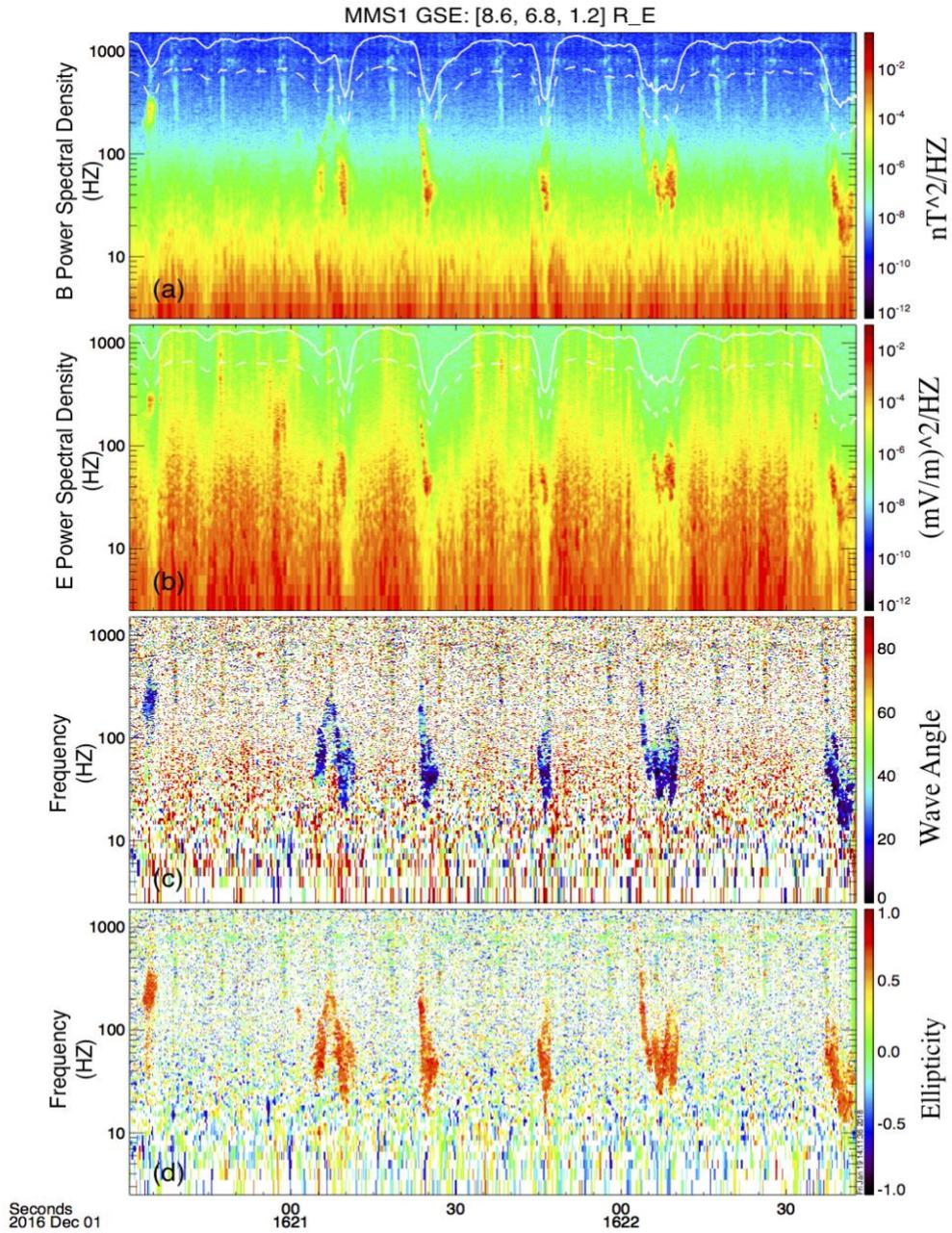

Figure 4: Wave polarization analysis of mirror instability event using Search Coil Magnetometer (SCM) instrument and FGM instrument. (a) Magnetic field power spectral density with white solid and dashed line representing electron gyrofrequency and half electron gyrofrequency, respectively, (b) electric field power spectral density, (c) wave angle, (d) ellipticity. The whistler waves are parallel propagating and right hand circularly polarized with the ellipticity of 1.

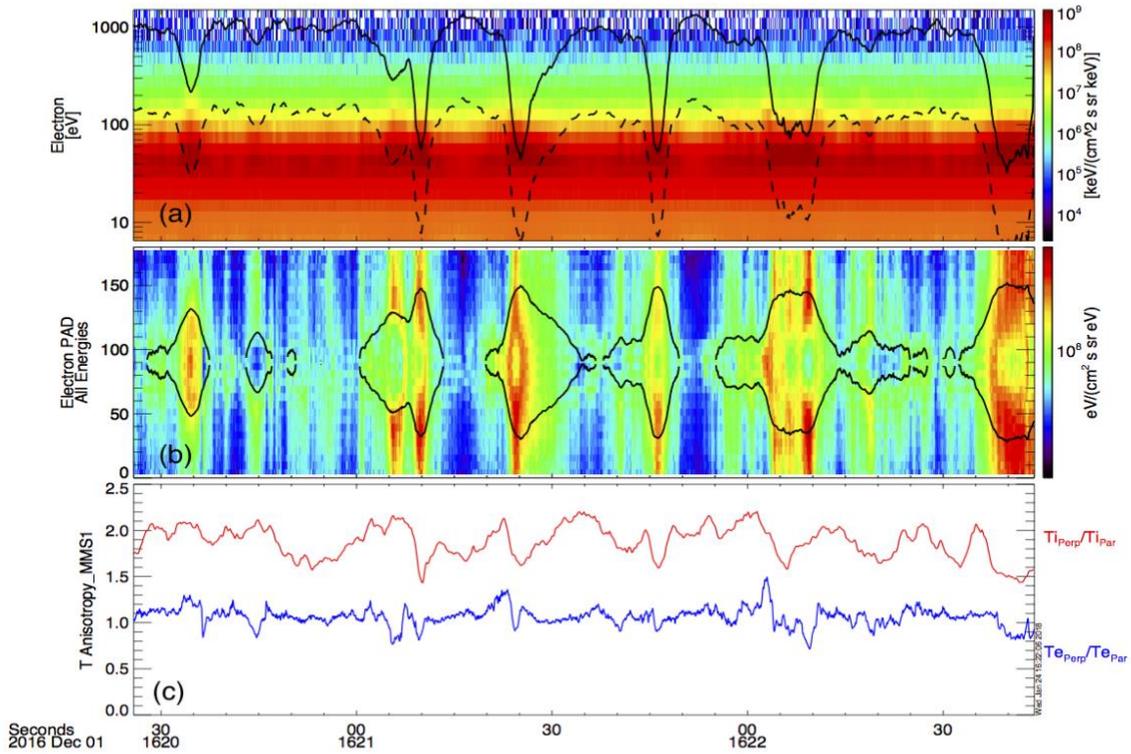

Figure 5: High resolution measurements with the FPI instrument of electron pitch angle distributions. (a) Electron energy spectrogram (the electron resonance energies are over-plotted by the black solid line for $\omega/\Omega_e = 0.12$ and black dashed line for $\omega/\Omega_e = 0.35$), (b) electron pitch angle distribution for all energies (over-plotted solid lines are the trapped-passing boundary), (c) proton and electron temperature anisotropies

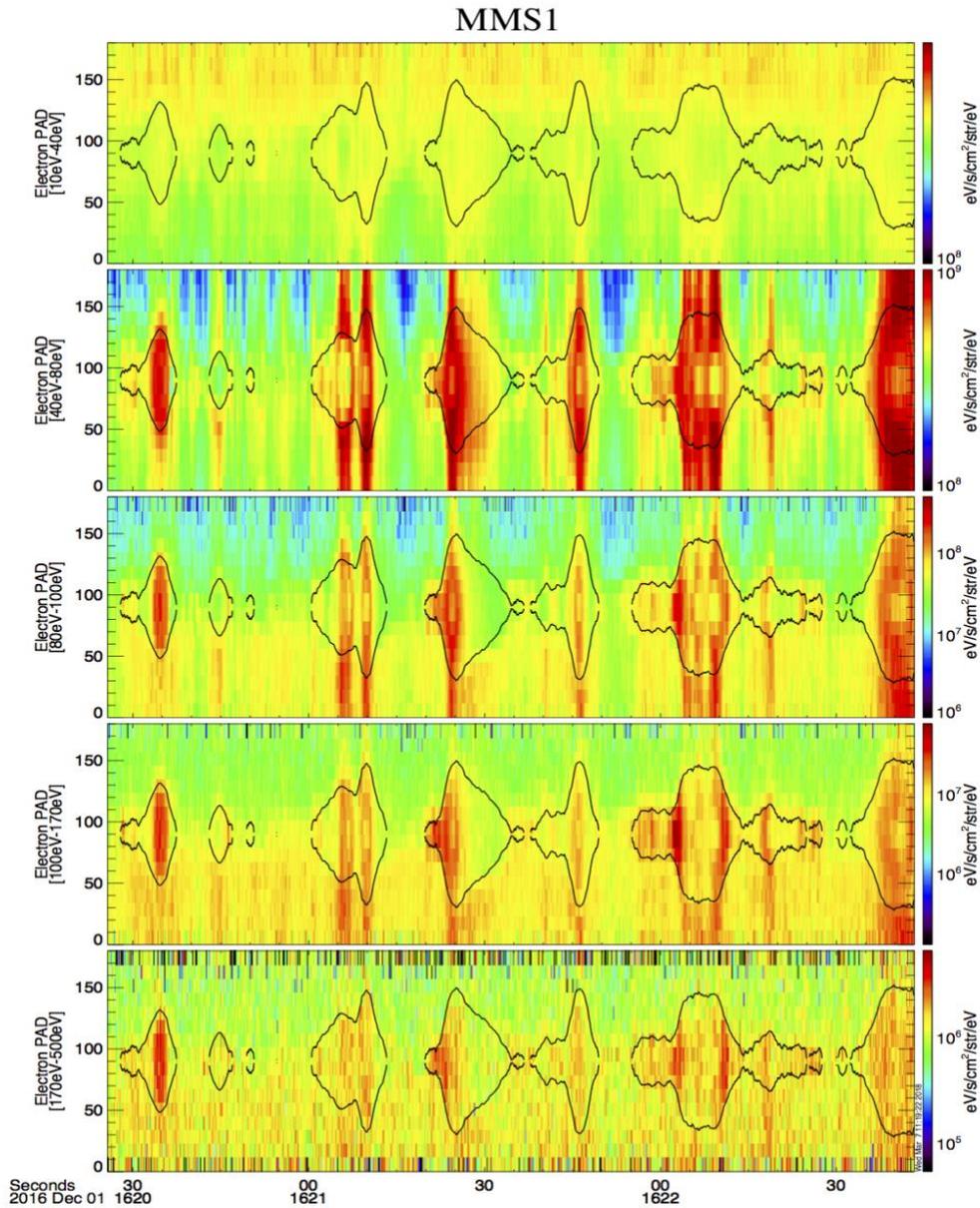

Figure 6: Pitch angle distributions (PAD) of electrons using the FPI burst data for different energy ranges. The color scale is different for each panel. The solid black lines represent the trapped-passing boundary. (a) Energy range 10-40eV (b) Energy range 40-80eV (c) Energy range 80-100eV (d) Energy range 100-170eV (e) Energy range 170-500eV. Electrons with energies above 40eV are trapped inside the shallow magnetic holes at 16:20:40 while the deeper magnetic holes after 16:21:00 have different behavior in pitch angle distribution depending on the energy range.